\documentclass[twocolumn,twoside,showpacs,preprintnumbers,superscriptaddress,amsmath,amssymb]{revtex4}
 \usepackage{cases}

 \usepackage{mathrsfs}
 \usepackage{amssymb}
 \usepackage[dvips]{graphicx}
 \usepackage{dcolumn}
 \usepackage{bm}
 \usepackage{subfigure}
 \usepackage{floatflt}
 \usepackage{float}
 \usepackage{rotating}
 \usepackage{multirow}
 \usepackage[bookmarksnumbered,bookmarksopen,colorlinks,citecolor=blue,linkcolor=blue]{hyperref} %
 \usepackage{booktabs}
 \usepackage{epsfig,graphics,psfrag,amsmath,amssymb}
 \usepackage{sidecap}

\begin{document}
 \preprint{00-000}

 \title{Dynamical and statistical description of  multifragmentation in heavy-ion collisions }

 \author{Lihua Mao}
 \affiliation{College of Physics and Technology,
 Guangxi Normal University, Guilin 541004, China}

 \author{Ning Wang}
 \email{wangning@gxnu.edu.cn}
 \affiliation{College of Physics and Technology,
 Guangxi Normal University, Guilin 541004, China}
 \affiliation{State Key Laboratory of Theoretical Physics,
 Institute of Theoretical Physics, Chinese Academy of Sciences, Beijing 100190, China}

 \author{Li Ou}
 \email{liou@gxnu.edu.cn}
 \affiliation{College of Physics and Technology, Guangxi Normal University,
 Guilin 541004, China}
 \affiliation{State Key Laboratory of Theoretical Physics,
 Institute of Theoretical Physics, Chinese Academy of Sciences, Beijing 100190, China}

 \date{\today}

 \begin{abstract}
 To explore the roles of dynamical model and statistical model in the description of
 multifragmentation in heavy-ion collisions at intermediate energies,
 the fragments charge distributions of $^{197}$Au+$^{197}$Au at 35 MeV/u
 are analyzed by using the hybrid model of improved quantum molecular dynamics (ImQMD) model
 plus the statistical model GEMINI.
 We find that, the ImQMD model can well describe the charge distributions of fragments produced
 in central and semi-central collisions.
 But for the peripheral collisions of Au+Au at 35 MeV/u,
 the statistical model is required at the end of the ImQMD simulations
 for the better description of the charge distribution of fragments.
 By using the hybrid model of ImQMD+GEMINI, the fragment charge distribution of
 Au+Au at 35 MeV/u can be reproduced reasonably well.
 The time evolution of the excitation energies of primary fragments is simultaneously investigated.
 \end{abstract}

 \maketitle



\section{Introduction}

 The heavy-ion nuclear reactions at intermediate and high energies provide a unique approach
 to study the nuclear equation of state at high densities in terrestrial laboratories,
 and therefore attracted much attentions both in the experimental
 and theoretical fields \cite{Shvedov10,Ardouin97,Colonna92,Seres89,Watson88,Planeta08,Staszel01,Wada00}.
 Because of the very short time scale in heavy-ion collisions,
 it is difficult to record the dynamical process of reactions by the experimental equipment.
 What the experiments can provide are the phase space information of
 various particles produced in the reactions.
 To know what happened during the reactions, especially before the system equilibrium,
 reliable dynamical approaches are crucial to extract physics from heavy-ion collision experiments.
 Many transport theories have been developed to explain the relevant experimental data.
 The models can be classified into two main categories:
 one-body theory and $n$-body theory.
 The popular one-body theories include the time dependent Hartree-Fock \cite{Bonche76,Golabek09,Umar11}
 and the Boltzmann-Vlasov approach\cite{Bertsch84,Li93,Ou11},
 e.g. IBUU, RBUU, pBUU, RVUU, GiBUU, et. al, in BV family.
 One of the popular $n$-body theories is the molecular dynamics (MD) approach,
 e.g. QMD, AMD, ImQMD, IQMD, UrQMD, CoMD, et. al, in MD family.
 With many efforts on the study of heavy-ion reactions,
 more and more observables are obtained and need to be explained by the transport model,
 the reliability of transport models faces a great challenge.
 One of the most challenging tasks for the $n$-body theory in nuclear physics
 is the description of the multifragmentation of heavy nuclei.
 The investigation of multifragmentation \cite{Bertsch88,Ono04,Hartnack98,Das05,Bondorf95,Gross97}
 is important for understanding the reaction mechanism in heavy-ion collisions.
 One-body theories are not proper tools for the investigation of
 many-body correlations, because the formation of fragments is beyond the scope of these models.
 Another kind of theory is the statistical approaches, i.e. statistical or thermodynamical models.
 Such as canonical thermodynamical model (CTM) \cite{Das05},
 statistical multifragmentation model (SMM) \cite{Bondorf95},
 GEMINI model \cite{Charity88,Charity01}, GEM2 \cite{Furihata00,Furihata02} model and HIVAP code \cite{Reidsdorf85}.
 The statistical approach is more appropriate to study the nuclei in equilibrium state.
 So the statistical approach is used to describe the spectator in peripheral heavy-ion collisions.

 The $n$-body theory can well describe the multifragmentation process
 in central and semi-central heavy-ion reactions at intermediate and high energies,
 in which the large density fluctuations caused by compression and expansion
 and high excitation energies of the system play a dominant role for the formation of fragments.
 However, the $n$-body theory fails to self-consistently describe the peripheral heavy-ion collisions
 and spallation reactions with heavy nuclei
 because in this kind of reactions the density fluctuations and excitation energies are relatively small
 and the fission of composite system or heavy fragments could not be neglected.
 So the statistical analysis at the end of dynamical simulations is necessary and important,
 since the transport models only contain classical correlations,
 which is insufficient to correctly describe evaporation (where realistic density of states are needed)
 or fission (where quantum fluctuations are essential).
 On the other hand, low energy reactions require a simulation with very long time.
 In practice it is a massive mission to simulate the reaction of very heavy system
 with very large time scale. The hybrid model by dynamical plus statistical model
 is a good compromise to study this kind of subject.

 In this paper, we use the improved quantum molecular dynamics (ImQMD) model
 plus the statistical model GEMINI to describe the multifragmentation process in heavy ion collisions.
 The paper is organized as follows.
 In sec. II, we briefly introduce the model we adopted.
 In sec. III, we present some calculations about the excitation energy
 and charge distribution of fragments in the reactions of $^{197}$Au+$^{197}$Au
 at incident energies of $E_{\rm{c.m.}}=$35 MeV/u. Finally a brief summary is given in Sec. IV.

\section{Theoretical approaches}

 In the ImQMD model \cite{Wang02}, as in the original QMD model \cite{Aichelin91},
 each nucleon is represented by a coherent state of a Gaussian wave packet.
 \begin{eqnarray} \label{wp}
  \phi_{i}(\bm{r})=\frac{1}{(2\pi\sigma_{r}^{2})^{3/4}}
  \exp\left[-\frac{(\bm{r}-\bm{r}_{i})^{2}}{4\sigma_{r}^{2}}
  +\frac{i}{\hbar}\bm{r}\cdot\bm{p}_{i}\right],
 \end{eqnarray}
 where $\bm{r}_i$ and $\bm{p}_i$ are the centers of the $i$th wave packet in the
 coordinate and momentum space, respectively.
 The one-body phase space distribution function is obtained by the Wigner transform of the wave function.
 The time evolution of $\bm{r}_i$ and $\bm{p}_i$ for each nucleon is governed by Hamiltonian equations of motion
 \begin{eqnarray}\label{Hequations}
 \dot{\bm{r}}_{i}=\frac{\partial H}{\partial\bm{p}_{i}}, \;\; \;
 \dot{\bm{p}}_{i}=-\frac{\partial H}{\partial\bm{r}_{i}},
 \end{eqnarray}
 where
 \begin{eqnarray}\label{Hamiton}
 H=T+U_{\rm{Coul}}+U_{\rm{loc}},
 \end{eqnarray}
 here, the kinetic energy $T=\sum\limits_{i}\frac{\bm{p}_{i}^{2}}{2m}$,
 $U_{\rm{Coul}}$ is the Coulomb energy,
 and the local potential energy $U_{\rm{loc}} =\int V_{\rm{loc}}[\rho(\bm{r})]d \bm{r}$.
 $V_{\rm{loc}}$ is the nuclear potential energy density functional that is obtained
 by the effective Skyrme interaction, which reads
 \begin{eqnarray}\label{Vloc}
 V_{\rm{loc}}&=&\frac{\alpha}{2}\frac{\rho^{2}}{\rho_{0}}+
 \frac{\beta}{\gamma+1}\frac{\rho^{\gamma+1}}{\rho_{0}^{\gamma}}+
 \frac{g_{\rm{sur}}}{2\rho_{0}}(\nabla \rho)^{2} \nonumber \\
 &&+\frac{C_{\rm{s}}}{2\rho_{0}}[\rho^{2}-\kappa_{\rm{s}}(\nabla\rho)^{2}]\delta^{2}
 +g_{\tau}\frac{\rho^{\eta+1}}{\rho_0^{\eta}}.
 \end{eqnarray}
 In this work, we adopt the parameter set IQ3 \cite{Zanganeh12} (see Table \ref{parameter}),
 which has been proposed and tested for describing the heavy-ion fusion reactions and the multifragmentation process,
 in the previous works \cite{Zanganeh12,Li13}.
 \begin{table*}
 \caption{\label{parameter}Parameter set IQ3 used in the ImQMD calculations.}
 \begin{tabular}{ccccccccccc}
 \hline\hline
 $\alpha $ & $\beta $ & $\gamma $ &$%
 g_{\rm{sur}}$ & $ g_{\tau }$ & $\eta $ & $C_{s}$ & $\kappa _{s}$ &
 $\rho_{0}$ & ~~$\sigma_0$~~ & ~~$\sigma_1$~~ \\
  (MeV) & (MeV) &  & (MeV~fm$^{2}$) & (MeV) &  & (MeV) & (fm$^{2}$) &
 (fm$^{-3}$) & (fm) & (fm) \\ \hline
  $-207$ & 138 & 7/6 & 18.0 & 14.0 & 5/3 &  32.0  & 0.08  & 0.165 & 0.94 & 0.018\\
 \hline\hline
 \end{tabular}
 \end{table*}

 At the end of the ImQMD calculations, clusters are recognized by a minimum spanning tree (MST) algorithm \cite{Aichelin91}
 widely used in the QMD calculations. In this method, the nucleons with relative momenta smaller
 than $P_{\rm{c}}$ and relative distances smaller than $R_{\rm{c}}$ are coalesced into the same cluster.
 In this work, $R_{\rm{c}}$ = 3.5 fm and $P_{\rm{c}}$ = 300 MeV/$c$ are adopted.
 Then the total energy of each excited cluster is calculated in its rest frame
 and its excitation energy are obtained by subtracting the corresponding ground state energy
 from the total energy of the excited cluster.
 The information of excited cluster is input into the statistical decay model GEMINI
 to perform statistical decay stage calculations.

 The GEMINI is based on the well-known sequential-binary-decay picture
 that the individual compound nuclei decay through sequential binary decays of all possible modes,
 from emission of nucleons and light particles through asymmetric
 to symmetric fission as well as the $\gamma$ emission,
 until the resulting products are unable to undergo any further decay.
 Fortran95 version (released on 2-May-2005) of GEMINI is used in this work.
 A very detailed description of the GEMINI can be found in Ref.\cite{Charity88,Charity01}
 and references therein.
 The GEMINI parameters are chosen as following:
 The level density is taken as Grimes case B modified form ({\it aden\_type}=$-$23),
 all asymmetric divisions is considered in fission mode ({\it imf\_option}=2),
 the particle with $Z\le 5$ is treated in light particle evaporation ({\it Z\_imf\_min}=5).
 The other parameters are taken as the default value given by example in GEMINI document.

\section{Results and discussion}

 As the most important input of GEMINI model, the excitation energy
 determines the decay process of the primary cluster formed in the dynamical stage.
 So the calculation of excitation energy should be checked carefully.
 We firstly prepare the two initial nuclei with their properties of the ground states being well described.
 The time evolutions of the binding energies
 and nuclear radii for a number of nuclei have been checked simultaneously.
 Only those individual nucleus which can remain stable for several
 thousands fm/$c$ are taken to be good initial nuclei,
 and then are applied in the simulation of the reaction process.

\begin{figure*}[htpb]
 \includegraphics[angle=0,width=1.0\textwidth]{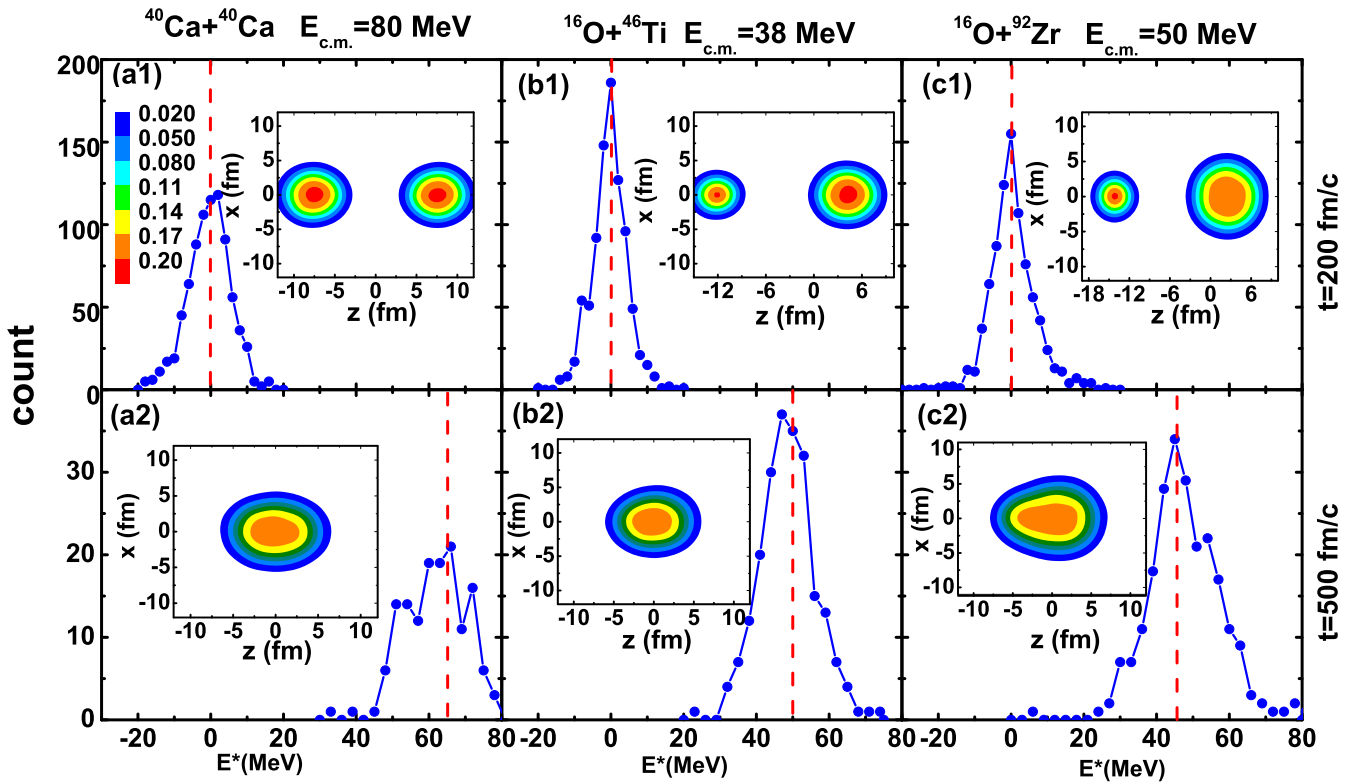}
 \caption{(Color online) Excitation energy distribution of the largest fragments for
 head-on collisions of $^{40}$Ca+$^{40}$Ca ($E_{\rm{c.m.}} =80$ MeV) (left panel),
 $^{16}$O+$^{46}$Ti ($E_{\rm{c.m.}} =38$ MeV) (middle panel),
 and $^{16}$O+$^{92}$Zr ($E_{\rm{c.m.}} =50$ MeV) (right panel),
 at 200 fm/$c$ (top panel) and 500 fm/$c$ (bottom panel), respectively.
 The contour plot of nuclear density distribution is also presented.}\label{fig1}
 \end{figure*}

 Figure \ref{fig1} presents the excitation energy distribution for the head-on collisions of
 $^{40}$Ca+$^{40}$Ca ($E_{\rm{c.m.}} =80$ MeV), $^{16}$O+$^{46}$Ti ($E_{\rm{c.m.}} =38$ MeV),
 and $^{16}$O+$^{92}$Zr ($E_{\rm{c.m.}} =50$ MeV), at 200 fm/$c$ and 500 fm/$c$, respectively.
 To check the procedure of the calculation of excitation energy in the ImQMD model,
 the corresponding excitation energy $ E_Q^{*}=E_{\rm{c.m.}}+Q_{\rm{gg}}$ of the compound nuclei
 in heavy-ion fusion reaction is also presented for comparison.
 Here $Q_{\rm gg}$ denotes the $Q$ value of the fusion reaction from ground state to ground state.
 $Q_{\rm{gg}}=-$14.3, 12.3, and $-3.9$ MeV for $^{40}$Ca+$^{40}$Ca,
 $^{16}$O+$^{46}$Ti and $^{16}$O+$^{92}$Zr, respectively.
 The corresponding values of $E_Q^{*}$ are 65.7, 50.3 and 46.1 MeV
 for the three reaction systems (see the dashed lines), respectively.
 One can see that at $t$ =200 fm/$c$, the projectile and target are well separated,
 so the excitation energies of target nuclei are distributed around zero. At $t$ =500 fm/$c$,
 the compound nuclei are generally formed and the peaks of the excitation energies distribution
 locate at around the corresponding $E_Q^{*}$.
 These tests indicate that the calculation results of excitation energies of primary fragments are reasonable.

\begin{figure*}[htpb]
 \includegraphics[angle=0,width=1.0\textwidth]{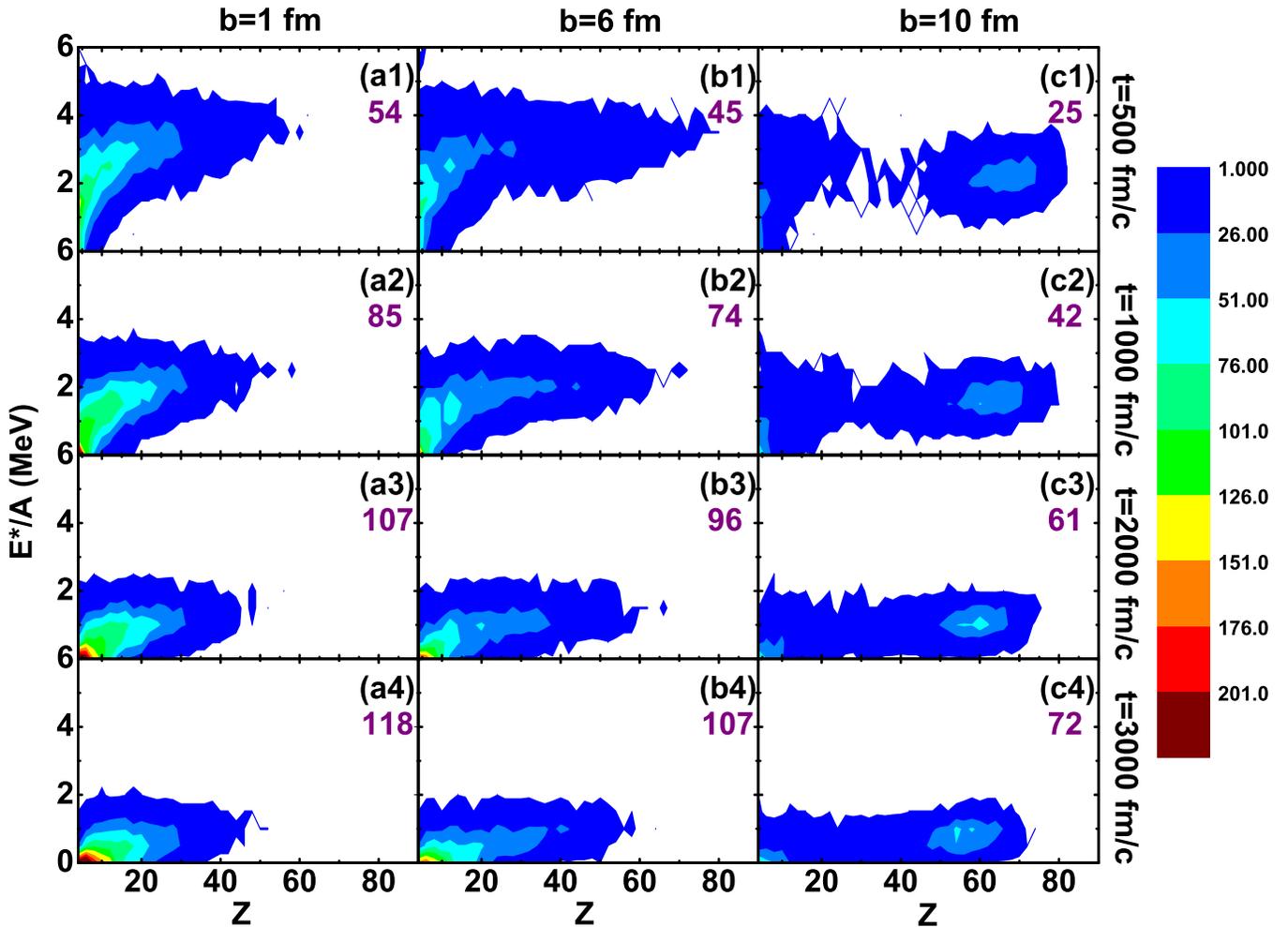}
 \caption{(Color online) Time evolutions of the excitation energies per nucleon of
 primary fragments with charge number $Z \ge 4$ for $^{197}$Au+$^{197}$Au at incident energy of $E_{\rm{c.m.}} =35$ MeV/u.
 The average numbers of free nucleons at each event are also presented in each sub-figures.}
 \label{fig2}
 \end{figure*}

 The reactions of $^{197}$Au+$^{197}$Au at incident energies of $E_{\rm{c.m.}} =35$ MeV/u
 are simulated with the ImQMD model to investigate the multifragmentation behavior of heavy-ion collision.
 Firstly, we check the time evolutions of excitation energies  per nucleon of primary fragments
 formed in reactions with impact parameters $b$=1, 6, and 10 fm, respectively, as presented in Fig. \ref{fig2}.
 One can get a direct impression that with time evolution, both
 the excitation energies and maximum fragments become small.
 In the case of central and semi-central collisions,
 the multifragmention process can be obviously observed.
 The collective incident kinetic energy has transferred to the excitation energy
 due to the collisions between two nuclei with same size.
 Then the excited fragments decay by emitting free nucleons and light charged particles.
 So the yield of fragment with charge $Z \le$ 10 increase with time evolution.
 In the case of peripheral collisions, after $t=$ 500 fm/$c$,
 there is only a few intermediate mass fragments (IMFs) yielded.
 The spectators with very high excitation energies are formed.
 With time evolution, the spectators decay by emitting free nucleons,
 and it only increases the yield of fragment with change around 60,
 but has a little contribution to the yield of IMF.
 Even until 3000 fm/$c$, there are still certain number of
 fragments with high excitation energies.
 It indicates that the statistical description is necessary for the peripheral collisions of this reaction.

\begin{figure*}[htpb]
 \includegraphics[angle=0,width=1.0\textwidth]{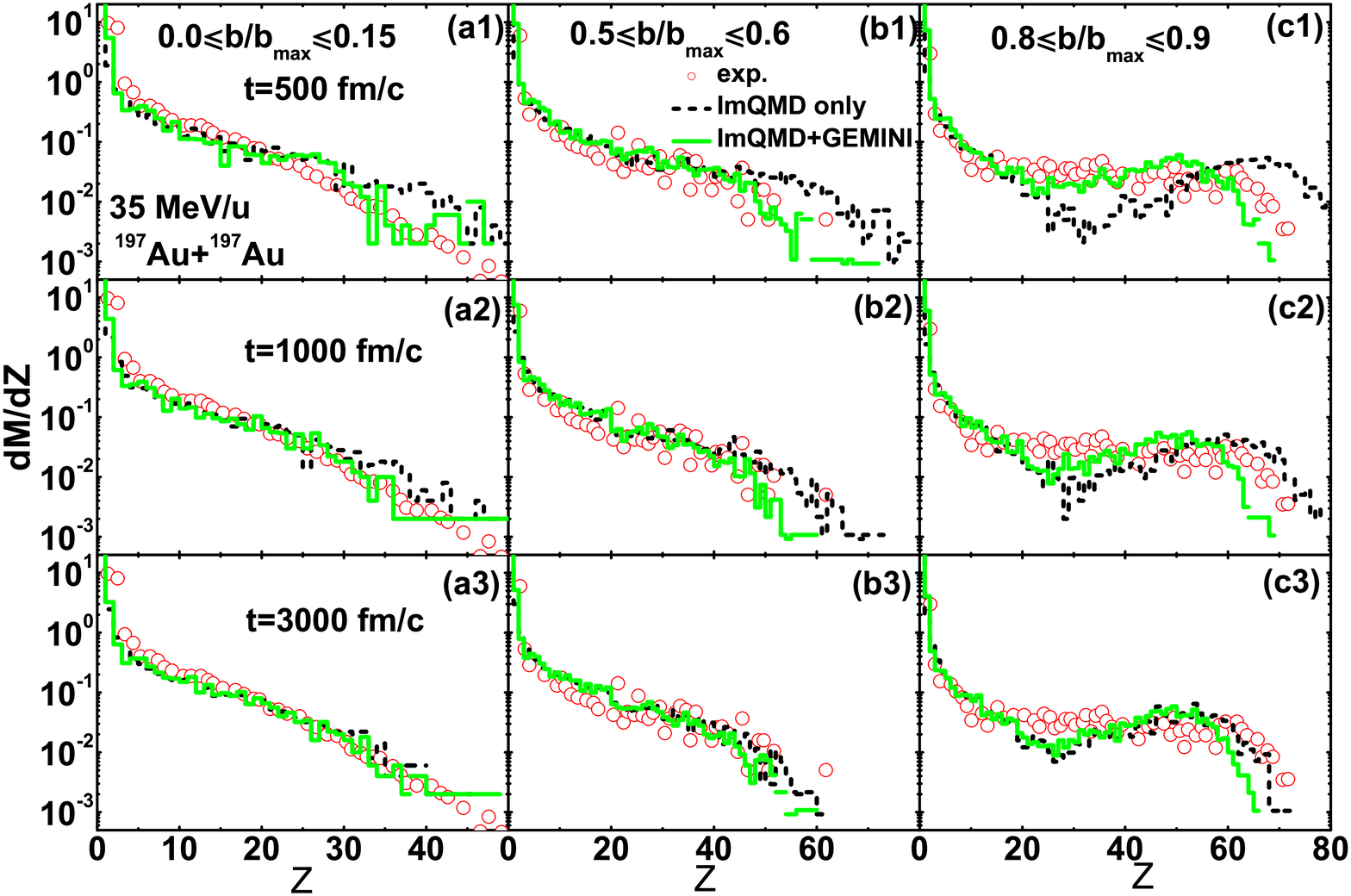}
 \caption{(Color online) Comparison between calculations and experimental data
 for charge distributions of fragments in $^{197}$Au+$^{197}$Au.
 The experimental data are taken from Ref. \cite{Lavaud02} for central collision,
 and Ref. \cite{Agostino99} for semi-central and peripheral collision, respectively.}
 \label{fig3}
 \end{figure*}

 By using the hybrid model of ImQMD+GEMINI, we study the charge distribution
 of fragments in the reactions of $^{197}$Au+$^{197}$Au at incident energies of $E_{\rm{c.m.}} =35$ MeV/u.
 The comparison between calculations and experimental data are presented in Fig. \ref{fig3}.
 It should note that the incident energy is 40 MeV/u for experimental data of central collision.
 We do not think there should be quite difference to the data of 35 MeV/u.
 According to the experiment setup,
 in the case of central collision, products in the center of mass angles between 0$^{\circ}$ and 90$^{\circ}$
 are chosen to compare with the experimental data \cite{Lavaud02}.
 In the semi-central and peripheral collision,
 products with charge up to the beam charge at $\theta_{\rm{lab}}$ from 3$^{\circ}$ to 23$^{\circ}$,
 and products with charge up to $Z$=20 covering the angular range from 23$^{\circ}$ to 160$^{\circ}$
 are chosen to compare with experimental data \cite{Agostino99}.
 One can see from the figure that, in the case of central and semi-central collision,
 the ImQMD model reproduces the experimental data reasonably well even without the statistical calculations.
 In the early stage of reactions, there are still some large fragments with high excitation energies.
 With time evolution, the results directly from the ImQMD model approach gradually to those from the ImQMD+GEMINI calculations,
 because the excitation energies of the fragments are exhausted by emitting particles.
 But in the case of peripheral collisions, the calculations without the statistical model obviously
 underestimate the yield of IMFs with $20 \le Z \le 40$.
 It needs a much longer time for the heavy fragments to undergo large deformation and breakup.
 Because of the absence of some Fermionic properties in the ImQMD model,
 some nucleons with high momentum will be emitted more easily and take away some excitation energies.
 With smaller and smaller excitation energy, the heavy fragments become difficult to breakup.
 One can also see that, even the simulation stops at 3000 fm/$c$,
 there is still obvious difference between the results with and without the statistical model being involved.
 It means that some fragments with high excitation energies will enter the secondary decay process.
 From the calculations, we find that the experimental data can be well reproduced
 with the ImQMD+GEMINI model at switch time $t$=500 fm/$c$.
 Finally what we want to mention is that, without any readjusting,
 our model can simultaneously reproduce the data
 taken from two different experiments with very different apparatus:
 INDRA+ALADIN for the central collisions, and MULTICS for the peripheral ones.
 It illustrates that our model has powerful prediction ability.
 We note that, in the present calculations, the angular momentum of fragments is not taken into account
 for the simplicity.
 In the fragmentation process, the emissions of nucleons and light particles
 could take away some angular momentum of heavy fragments.
 The neglect of angular momentum might be at the origin of
 the (small) discrepancy with experimental data which is still observed
 for the most peripheral collisions. The investigate on the influence of
 angular momentum of fragments is under way.
\section{Summary}

 The multifragmentation in the intermediate heavy-ion collisions
 has been investigated by using the hybrid model of Improved Quantum Molecular Dynamics model plus the statistical model GEMINI.
 The excitation energy of fragments in heavy-ion fusion reactions is firstly tested with the ImQMD model,
 and the calculation results look quite reasonable. The roles of dynamical model and statistical model
 in the description of the reaction mechanism
 are investigated by analyzing on the fragments charge distribution of 35 MeV/u $^{197}$Au+$^{197}$Au
 at various impact parameters. We find that, in the case of central and semi-central collisions,
 the calculations without the statistical calculations are in good agreement with the measured fragments charge distribution.
 The statistical process with nucleons and light charge particles emission can be partially represented by dynamical model.
 However, in the case of peripheral collisions induced by heavy nuclei,
 the statistical model is still necessary to describe the multifragmentation.
 Otherwise the IMFs yields will be underestimated due to the absence of quantum correlations in the ImQMD model.
 By using the hybrid model of ImQMD+GEMINI, the charge distribution of fragments
 in Au+Au at 35 MeV/u can be reproduced reasonably well for both central and peripheral collisions.

 \begin{acknowledgments}
 This work has been supported by the National Natural Science Foundation of China under Grant Nos.
 11005022, 
 11365004, 
 11365005, 
 11422548, 
 11275052, 
 11475262, 
 11475004 
 and the Open Project Program of State Key Laboratory of Theoretical Physics, Institute
 of Theoretical Physics, Chinese Academy of Sciences, China (No. Y4KF041CJ1).
 \end{acknowledgments}





\begin{thebibliography}{99}


 \bibitem{Shvedov10} L. Shvedov, M. Colonna, and M. Di Toro, Phys. Rev. C. \textbf{81}, 054605 (2010).

 \bibitem{Ardouin97} D. Ardouin, Int. J. Mod. Phys. E. \textbf{6}, 391 (1997).

 \bibitem{Colonna92} M. Colonna, M. Di Toro, V. Latora, and N. Colonna, Nucl. Phys. A \textbf{545}, 111 (1992).

 \bibitem{Seres89} Z. Seres, F. De'ak, A. Kiss, and G. Caskey \textit{et al.}, Nucl. Phys. A \textbf{492}, 315 (1989).

 \bibitem{Watson88} R. L. Watson, R. J. Maurer, B. B. Bandong, and C. Can, Lect. Notes Phys. \textbf{294}, 382 (1988).

 \bibitem{Planeta08} R. Planeta, F. Amorini, and A. Anzalone \textit{et al.}, Phys. Rev. C \textbf{77}, 014610 (2008).

 \bibitem{Staszel01} P. Staszel, Z. Majka, and L. G. Sobotka \textit{et al.}, Phys. Rev. C \textbf{63}, 064610 (2001).

 \bibitem{Wada00} R. Wada, K.Hagel, J. Cibor \textit{et al.}, Phys. Rev. C \textbf{62}, 034601 (2000).

 \bibitem{Bonche76} P. Bonche, S. Koonin and J. W. Negele, Phys. Rev. C \textbf{13}, 1226 (1976).

 \bibitem{Golabek09} C. Golabek, and C. Simenel, Phys. Rev. Lett. \textbf {103}, 042701 (2009).

 \bibitem{Umar11} A. S. Umar, V. E. Oberacker, J. A. Maruhn,
 and P.-G. Reinhard, Euro. Phys. J. Web. Conf. \textbf{17}, 9001 (2011).

 \bibitem{Bertsch84} G. F. Bertsch, H. Kruse, and S. DasGupta, Phys. Rev. 29, 673 (1984).

 \bibitem{Li93} B.-A. Li, and D. H. E. Gross, Nucl. Phys. A \textbf{554}, 257 (1993).

 \bibitem{Ou11} L. Ou, and B.-A. Li, Phy. Rev. C \textbf{84}, 064605 (2011).

 \bibitem{Bertsch88} G. F. Bertsch, and S. Das. Gupta, Phys. Rep. \textbf{160}, 189 (1988).

 \bibitem{Ono04} A. Ono, and H. Horiuchi, Prog. Part. Nucl. Phys. \textbf{53}, 501 (2004).

 \bibitem{Hartnack98} C. Hartnack, \textit{et al.}, Eur. Phys. J. A \textbf{1}, 151 (1998).

 \bibitem{Das05} C. B. Das, \textit{et al.}, Phys. Rep. \textbf{406}, 1 (2005).

 \bibitem{Bondorf95} J. P. Bondorf, \textit{et al.}, Phys. Rep. \textbf{257}, 133 (1995).

 \bibitem{Gross97} D. H. Gross, Phys. Rep. \textbf {279}, 119 (1997).

 \bibitem{Charity88} R. J. Charity, M. A. McMahan, G. J. Wozniak \textit{et al.}, Nucl. Phys. A \textbf{483}, 371 (1988).

 \bibitem{Charity01} R. J. Charity, L. G. Sobotka, Y. E. Masri \textit{et al.}, Phys. Rev. C \textbf{63}, 024611 (2001).

 \bibitem{Furihata00} S. Furihata, Nucl. Instrum. Methods Phys. Res. B \textbf{171}, 251 (2000).

 \bibitem{Furihata02} S. Furihata and T. Nakamura,  J. Nucl. Sci. Technol. Suppl. \textbf{2}, 758 (2002).

 \bibitem{Reidsdorf85} W. Reisdorf \textit{et al.}, Nucl. Phys. A \textbf{444}, 154 (1985).

 \bibitem{Wang02} N. Wang, Z. X. Li, and X. Z. Wu, Phy. Rev. C \textbf{65}, 064608 (2002).

 \bibitem{Aichelin91} J. Aichelin, Phys. Rep. \textbf{202}, 233 (1991).

 \bibitem{Zanganeh12} V. Zanganeh, N. Wang, and O. N. Ghodsi, Phys. Rev. C \textbf{85}, 034601(2012).

 \bibitem{Li13} C. Li, J. L. Tian, L. Ou, and N. Wang, Phys. Rev. C \textbf{87}, 064615 (2013).

 \bibitem{Agostino99} M. D'Agostino, A. S. Botvina, M. Bruno,
 \textit{et al.}, Nucl. Phys. A \textbf{650}, 329 (1999).

 \bibitem{Lavaud02} F. Lavaud, E. Plagnol, G. Auger \textit{et al.}, AIP Conf. Proc. 610, 716 (2002).

 \end{thebibliography}
\end{document}